%% file: main_ICCS.tex
\documentclass[runningheads]{llncs}
\usepackage[english]{babel}           
\usepackage[utf8]{inputenc}           
\usepackage{authblk}

\usepackage{graphicx}    
\usepackage{color}       
\usepackage{anysize}     
\usepackage{multicol}    
\usepackage{bm}          
\usepackage{textcomp} 
\usepackage{eurosym}  
\usepackage{amsmath,amsfonts}
\usepackage{lineno}     
\usepackage{fancyhdr}
\usepackage{float}
\usepackage[hidelinks]{hyperref}
\usepackage[normalem]{ulem}
\usepackage{multirow}
\usepackage[table,xcdraw]{xcolor}
\usepackage{longtable}
\usepackage{subfigure}
\usepackage{mathtools}
\usepackage{braket}
\usepackage{qcircuit}
\usepackage[nameinlink, capitalise]{cleveref}

\marginsize{3cm}{3cm}{2cm}{2cm}
\usepackage[colorinlistoftodos]{todonotes}
\usepackage{xcolor}
\definecolor{darkgreen}{rgb}{0.0, 0.4, 0.0}
\usepackage{braket}

\title{Noise robustness of a multiparty quantum summation protocol}
\author{Ant{\'o}n Rodríguez-Otero \and Niels M. P. Neumann \and Ward van der Schoot \and Robert Wezeman}
\authorrunning{A. Rodríguez-Otero et al.}

\institute{The Netherlands Organisation for Applied Scientific Research, The Netherlands, The Hague}

\begin{document}

\maketitle
\begin{abstract}
Connecting quantum computers to a quantum network opens a wide array of new applications, such as securely performing computations on distributed data sets. 
Near-term quantum networks are noisy, however, and hence correctness and security of protocols are not guaranteed. 
To study the impact of noise, we consider a multiparty summation protocol with imperfect shared entangled states.
We study analytically the impact of both depolarising and dephasing noise on this protocol and the noise patterns arising in the probability distributions. 
We conclude by eliminating the need for a trusted third party in the protocol using Shamir's secret sharing. 
\keywords{Distributed Quantum Computing \and Noisy Quantum Communication \and Multi-Party Computation \and Shamir Secret Sharing}

\end{abstract}

\input{paper}
\bibliographystyle{splncs04}
\bibliography{main_ICCS}
\end{document}

%% file: paper.tex
\section{Introduction}
Quantum computing is an emerging field where advances are made on the hardware-side, software-side, as well as applications. 
Many companies and universities are working on building better quantum hardware with more resources of better quality. 
At the same time, new algorithms are being discovered, and these new quantum algorithms are applied in various new settings. 

The theoretical speedup quantum computers offer for various problems discerns them from classical alternatives.
Amongst these are some of the most complicated problems encountered in every-day life.
Examples where quantum computers outperform classical alternatives include breaking certain asymmetric encryption protocols~\cite{Shor_1994,Shor_1997}, developing new materials and personalised medicines~\cite{Fedorov2021}, and solving complex systems of linear equations~\cite{HHL:2009}.

Another aspect at which quantum computers distinguish themselves from classical alternatives, is the security of a quantum state: 
Opposed to classical information, in general, quantum information cannot be read out or copied faithfully. 
Reading out a quantum state destroys the state irrevocably and loses information, whereas trying to copy a quantum state leaves the state and its copy entangled, and operations performed on an entangled copy differ from those applied to the original unentangled state.
Because of this, sharing information via quantum states is secure. 
This idea underlies the field of quantum communication and its subfield quantum key distribution. 

Combining quantum computing with quantum communication joins the best of both worlds: 
by using quantum communication between different quantum computers, these devices can collaboratively solve larger problems, while the information shared between the devices remains secure. 
This field is called \textit{Distributed Quantum Computing} (DQC). 

Distributed quantum computing entails the collaborative execution of quantum algorithms using multiple quantum devices. 
Distributed computations can occur at various levels: 
for example, the devices may independently run their own quantum circuits, after which the outputs are combined to obtain the final results.
Alternatively, the devices may cooperate intricately through quantum communication to execute a single overall circuit. 
This study concentrates on the latter scenario, specifically exploring the execution of a distributed quantum addition circuit.

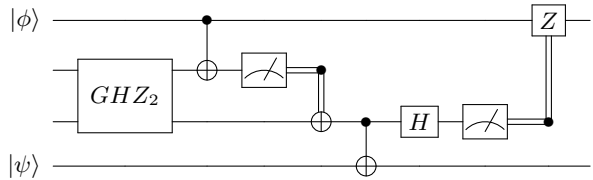
\begin{figure}[ht!]
	\centering
	${\Qcircuit @C=1em @R=0.7em {
	    \lstick{\ket{\phi}} & \qw & \ctrl{1} & \qw & \qw & \qw & \qw & \qw & \gate{Z} & \qw & \\
		& \ghost{GHZ_2} & \targ & \meter & \control\cw\cwx[1] &&&& \\
		& \multigate{-1}{GHZ_2} & \qw & \qw & \targ & \ctrl{1} & \gate{H} & \meter & \control\cw\cwx[-2] \\
		\lstick{\ket{\psi}} & \qw & \qw & \qw & \qw & \targ & \qw & \qw & \qw & \qw & 
	}}$
	\caption{Quantum circuit for a non-local CNOT-gate between $\ket{\phi}$ and $\ket{\psi}$ located on different devices that share a GHZ-state. Double lines denote classical information.}
	\label{fig:dist_cnot_circ}
\end{figure}

The key challenge in this form of distributed quantum computing, is the application of non-local multi-qubit gates.
As any multi-qubit gate can be decomposed into CNOT gates with additional local one-qubit gates~\cite{Barenco:1995}, it suffices to implement the CNOT-gates in a non-local fashion.
Figure~\ref{fig:dist_cnot_circ} shows an implementation of a non-local CNOT gate between the states $\ket{\phi}$ and $\ket{\psi}$ located on different devices. 
The protocol requires shared entanglement between the two devices in the form of a $GHZ_2$ state.
The procedure involves two gates conditioned by a measurement outcome obtained by the other party.
Classical communication is thus needed between the parties in order to implement the non-local CNOT.
By replacing every CNOT acting on qubits from different devices by a non-local CNOT, any quantum circuit can be run in a distributed manner. 

Eisert et al. gave the first description of how to perform operations between different quantum devices through the use of quantum communication~\cite{Eisert:2000}. 
Later, this work was extended and a distributed version of Shor's algorithm was theorised~\cite{Yimsiriwattana2004,Yimsiriwattana_shor2004}. 
Distributed quantum computing works by transforming traditional quantum algorithms to their distributed version. 
In these distributed versions, operations performed between qubits located on different devices are called non-local and are replaced by a non-local quantum gate established using shared entangled states. 
In comparison, operations between qubits on the same device are called local, and are unchanged.
These three works consider all operations, both local and non-local, to be perfect. 
Beals et al. later proved that distributing an algorithm over different resources incurs only a small overhead in the cost~\cite{Beals2013}. 
Hence, when programming quantum algorithms on a higher level, the underlying structure of the hardware, local or distributed, has only a marginal effect. 

Follow-up work mainly focused on applications run using a distributed quantum network~\cite{DiAdamo:2021}, or on how to best implement a distributed quantum computer network~\cite{Gyongyosi2021,Caleffi2022}.
One aspect to take into account in these distributed networks is the robustness against noise, as current hardware is noisy and will remain so for the foreseeable future. 
It is therefore interesting to consider the effect of imperfect operations in such distributed settings. 
A first work on this topic computed the fidelity of a distributed and imperfect quantum phase estimation algorithm, when distributed over a varying number of devices~\cite{Neumann:2020}. 
Another example is the work by Khabiboulline et al. where a secure quantum voting protocol is presented~\cite{Khabiboulline}.

In this work, we extend this line of research by considering imperfect non-local operations as well, but applied to 
the distributed quantum summation protocol~\cite{Neumann2022}, which extends the algorithm proposed by Draper~\cite{Draper2000} and later improved by Ruiz-Perez and Garcia-Escartin~\cite{Ruiz-Perez2017}. 
The quantum summation algorithm uses the Quantum Fourier Transform to map the states to their phase state representation. 
In the phase space, addition corresponds to specific controlled phase gates.

In this protocol, we consider different parties which aim to compute the sum of their inputs, without revealing these inputs. 
Each party has access to a local quantum computer, which can generate shared entangled states with other devices. 
In practice, quantum hardware remains noisy and it is necessary to consider decoherence effects when developing applications.
Currently, the fidelity of state teleportation between non neighbouring nodes is around $0.7$~\cite{Hermans_2022} while the fidelity of quantum operations on quantum devices is around $0.95$ to $0.99$~\cite{li2023error}.
For this reason, we omit in this work the effect of imperfect quantum operations, and focus on the impact of an imperfect quantum network links. 
Concretely, we consider how dephasing and depolarising noise on the shared entangled states affects the output fidelity of this distributed summation protocol. 

We also extend this line of research by combining it with a primitive from cryptography called \textit{Shamir Secret Sharing}.
In earlier works, multiparty protocols are considered with the use of a central server party which is trusted by everyone.
This is not a realistic assumption in practical use cases.
In this work, we show how the considered multiparty summation protocol can be extended to a setting without the requirement of a trusted server party.
We show that the protocol yields the same output as the original protocol, while none of the parties learns inputs from other parties.

Section~\ref{sec:prelim} explains the multiparty summation protocol and the two considered noise models. 
Next, Section~\ref{sec:numerical_results} presents the results of simulations for both noise models.
Section~\ref{sec:analytical_study} contains an analytical study of the noise patterns and the periodicity therein.
Afterwards, Section~\ref{sec:protocol_modifications} details the extension of the protocol to a version without the need of a trusted server party.
Section~\ref{sec:conclusions} concludes with a summary and an outlook to future distributed quantum computing work.

\section{Preliminaries}\label{sec:prelim}
\subsection{Distributed Quantum Computing}
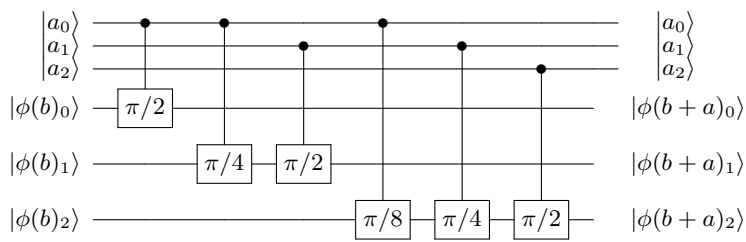
\begin{figure}[H]
	\centering
	\begin{minipage}[c]{0.5\textwidth}
		\centering
		{\Qcircuit @C=1em @R=0.7em {
			\lstick{\ket{a_0}} & \ctrl{3} & \ctrl{4} & \qw & \ctrl{5} & \qw & \qw & \qw & \qw & \rstick{\ket{a_0}} \\
			\lstick{\ket{a_1}} & \qw & \qw & \ctrl{3} & \qw & \ctrl{4} & \qw & \qw & \qw & \rstick{\ket{a_1}} \\
			\lstick{\ket{a_2}} & \qw & \qw & \qw & \qw & \qw & \ctrl{3} & \qw & \qw & \rstick{\ket{a_2}} \\
			\lstick{\ket{\phi(b)_0}} & \gate{\pi/2} & \qw & \qw & \qw & \qw & \qw & \qw & \rstick{\ket{\phi(b+a)_0}} \\
			\lstick{\ket{\phi(b)_1}} & \qw & \gate{\pi/4} & \gate{\pi/2} & \qw & \qw & \qw & \qw & \rstick{\ket{\phi(b+a)_1}} \\
			\lstick{\ket{\phi(b)_2}} & \qw & \qw & \qw & \gate{\pi/8} & \gate{\pi/4} & \gate{\pi/2} & \qw & \rstick{\ket{\phi(b+a)_2}}
		}}
	\end{minipage}
	\caption{The addition part of a quantum summation circuit. 
        The state $\ket{\phi(b)_j}$ represents the $j$-th bit of the Quantum Fourier transform of $b$. 
        The blocks represent controlled-$R_Z$ gates, where the argument shown is the angle used by the $R_Z$-gate. 
        An inverse quantum Fourier transform on the second half of the register gives the quantum state $\ket{a}\ket{b+a}$.}
	\label{fig:addition_fourier_state}
\end{figure}

We start by describing the non-distributed version of the quantum summation protocol, after which we explain the distributed version from \cite{Neumann2022}. 
Suppose we have two integers $a,b<2^n$ that we wish to add and that we have their corresponding quantum states $\ket{a}$ and $\ket{b}$ that are their binary representation using $n$ qubits.
The protocol first applies the quantum Fourier transform of size $n$, denoted by $QFT_n$ to $\ket{b}$. This yields the phase state representation of $b$, given by $\ket{\phi(b)}$. 
Then, applying phase gates to the qubits of $\ket{\phi(b)}$ controlled by the qubits of $\ket{a}$ gives the quantum state $\ket{\phi(a+b)}$. 
After applying an inverse quantum Fourier transform, the state $\ket{a+b}$, describing the binary representation of $a+b$, is obtained.
Figure~\ref{fig:addition_fourier_state} gives a schematic overview of the required phase gates of this protocol for $n=3$ qubits.
We omitted the two Fourier transforms. 

\begin{figure}[th]
	\centering
	\begin{minipage}[c]{0.5\textwidth}
		\centering
		{\Qcircuit @C=0.5em @R=0.45em {
			\lstick{s:\ket{0}} & \multigate{2}{QFT} & \ctrl{3} & \qw & \qw & \qw & \ctrl{3} & \qw & \qw & \multigate{2}{QFT^{-1}} & \qw & \\
			\lstick{s:\ket{0}} & \ghost{QFT} & \qw & \ctrl{3} & \qw & \qw & \qw & \ctrl{3} & \qw & \ghost{QFT^{-1}} & \qw & \\
			\lstick{s:\ket{0}} & \ghost{QFT} & \qw & \qw & \ctrl{3} & \qw & \qw & \qw & \ctrl{3} & \ghost{QFT^{-1}} & \qw & \\
			\lstick{p_1:\ket{0}} & \qw & \targ\qwx[3] & \qw & \qw & \multigate{2}{\text{Add } x^1} & \targ\qwx[3] & \qw & \qw & \qw & \qw & \rstick{\ket{0}} \\
			\lstick{p_1:\ket{0}} & \qw & \qw & \targ\qwx[3] & \qw & \ghost{\text{Add } x^1} & \qw & \targ\qwx[3] & \qw & \qw & \qw & \rstick{\ket{0}} \\
			\lstick{p_1:\ket{0}} & \qw & \qw & \qw & \targ\qwx[3] & \ghost{\text{Add } x^1} & \qw & \qw & \targ\qwx[3] & \qw & \qw & \rstick{\ket{0}} \\
			\lstick{p_2:\ket{0}} & \qw & \targ & \qw & \qw & \multigate{2}{\text{Add } x^2} & \targ & \qw & \qw & \qw & \qw & \rstick{\ket{0}} \\
			\lstick{p_2:\ket{0}} & \qw & \qw & \targ & \qw & \ghost{\text{Add } x^2} & \qw & \targ & \qw & \qw & \qw & \rstick{\ket{0}} \\
			\lstick{p_2:\ket{0}} & \qw & \qw & \qw & \targ & \ghost{\text{Add } x^2} & \qw & \qw & \targ & \qw & \qw & \rstick{\ket{0}} \\
		}}
	\end{minipage}
	\caption{Example of \textbf{DQA}~\cite{Neumann2022}.
        A server party adds integers from two computing parties. 
        The blocks $\text{Add }x^k$ denote the phase-gates needed to add integer $x^k$, similar to Figure~\ref{fig:addition_fourier_state}. 
        The final quantum state in the first register is $\ket{x^1 + x^2}$.}
	\label{fig:addition:method_1}
\end{figure}
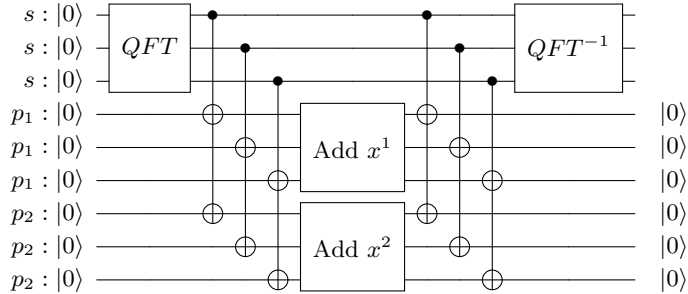

This summation protocol can be easily extended to allow a server party to do the addition of $k$ different numbers held by $k$ different computing parties. 
Figure~\ref{fig:addition:method_1} showcases the extended protocol for two computing parties with an additional server party. 
The server party holds the result at the end of the protocol. 
Note that the phase gates applied by different parties commute and hence, every party can apply their local phase gates simultaneously. 

The above multiparty protocol translates to a non-local protocol by replacing every CNOT gate by a non-local CNOT gate.
The resulting protocol is called the \emph{DISTRIBUTED-QFT-ADDER} (\textbf{DQA}). 
Multiple implementations for the CNOT-gates exist, some of which even allow simultaneous implementation of the phase gates by all parties~\cite{Neumann2022}.

\subsection{Noise models}
The quantum network distributes entangled $GHZ$ states between the server node and the different party nodes. 
The presence of noise translates to imperfect entanglement between the nodes. 

Current state-of-the-art protocols for entanglement generation between nodes of a quantum network are heralded, which allows to deterministically know whether the entanglement distribution process succeeded. 
Typically, heralding work in experiments via a photon measurement such that entanglement is established if and only if a photon has been detected.
We therefore disregard lossy quantum channels by assuming that the distribution is done in a heralded way.

In this study, we consider two types of noise in the quantum links, namely dephasing and depolarising noise.
These types of noise arise often in physical implementations and current software packages allow for easy simulation of these noise type, whilst the analytic study remains feasible at the same time. 
The respective noise channels are applied to the quantum links by applying them to all qubits in the $GHZ$ state independently. 

\subsubsection{Noise model A: Dephasing channel}
A dephasing channel is a \textit{completely positive and trace-preserving} (CPTP) map that represents the decay of the quantum phase of a system, that is, the off-diagonal elements of the density matrix. 
A one-qubit dephasing channel $\varepsilon_{depha}$ is usually represented by the map 
\begin{equation}
    \varepsilon_{depha}:\rho \mapsto \left(1-\frac{p}{2}\right)\rho + \frac{p}{2}\mathbf{Z}\rho\mathbf{Z}
    \label{eq:noise_dephasing}
\end{equation}
which performs a phase flip with probability $p/2$.
Writing the matrix representation of this channel we see, indeed, that it corresponds to a phase damping process:
\begin{equation}
    \rho = \begin{pmatrix}
        \rho_{00} & \rho_{01} \\
        \rho_{10} & \rho_{11}
    \end{pmatrix} \rightarrow
    \varepsilon_{depha}(\rho) = \begin{pmatrix}
        \rho_{00} & (1-p)\rho_{01} \\
        (1-p)\rho_{10} & \rho_{11}
    \end{pmatrix}.
\end{equation}

Dephasing errors arise in fiber optic links due to the birefringence phenomenon, which is associated with changes in the refractive index for different polarisations or different regions of the material~\cite{Wu_2013}. 
In fact, using a special polarisation maintaining optical fiber, the decoherence in these links can be completely described via dephasing processes~\cite{Jin-Shi2016}. 
Assuming heralded entanglement distribution, we can ignore the high attenuation arising in these links. 
Our focus is the impact of noise on the fidelity of the protocol, so we omit the entanglement generation rate and the lower transmission rates. 

\subsubsection{Noise model B: Depolarising channel}
The depolarising channel is usually seen as the quantum equivalent of white noise. 
Depolarising channels model processes that completely scramble the starting state with some probability. 
As a result, both quantum and classical information is lost. 
Given a valid $n$-qubit quantum state $\rho$, an $n$-qubit depolarising channel $\varepsilon_{depol}$ can be written as
\begin{equation}
    \varepsilon_{depol}:\rho \mapsto (1-p)\rho + \frac{p}{d}\mathbf{I},
\end{equation}
where $d=2^n$ is the dimension of the Hilbert space $\rho$ lives in.

Depolarising channels can, amongst other things, model the misalignment of reference frames between the nodes in a quantum network~\cite{safranek_2015}. 
Moreover, depolarising channels can also model what happens if heralding fails, for instance when the detector wrongfully measures a photon. 
Such an event is called a dark count and leads to reading out an empty quantum memory~\cite{vanDam2022}. 

\section{Simulations with noise}\label{sec:numerical_results}
We simulated the quantum summation protocol \textbf{DQA}~\cite{Neumann2022} and included the noise models discussed above to see how well the protocols perform in noisy settings. 
By adding dephasing or depolarising noise to the protocol, we expect incorrect outcomes found by the server party at the end of the protocol.
Therefore, we report the results as probability distributions in histograms or polar plots. 

We implemented both noise models using Qiskit~\cite{Qiskit} where we applied the noise models only to the quantum links, the part where the $GHZ$-states are generated.
We implemented these noise models by inserting local noisy identity gates at the end of the $GHZ$ generation block.
We ran experiments for varying number of parties and inputs, as well as noise levels. 
Each simulation consists of $9{,}000$ independent runs of the circuit\footnote{The implementations are available upon reasonable request to the authors}.

\subsection{Dephasing noise}
We first consider the impact of the dephasing noise by analysing four parties, each with input $1$ and a dephasing noise of $p/2=0.07$ for each of the four quantum links. 
Figure~\ref{fig:sum_1,1,1,1_histogram_p_07_dephasing} shows the resulting probability distribution in blue and the ideal (noiseless) probability distribution in red. 
\begin{figure}[H]
    \centering
    \includegraphics[width=0.9\textwidth]{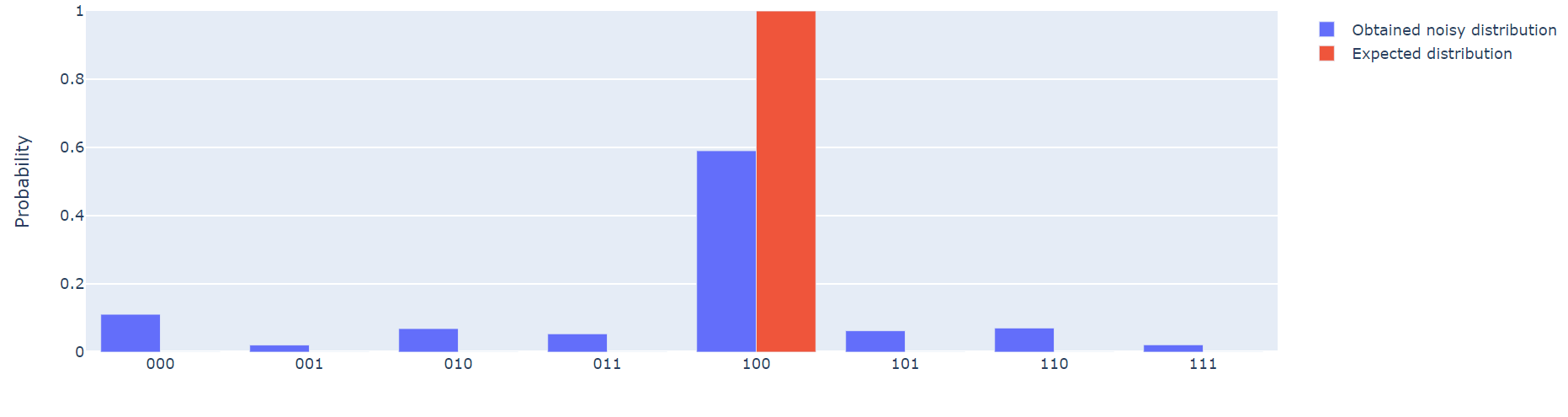}    
    \caption{Distribution of outcomes for the case of four parties inputting $1$, and a dephasing noise of $p/2=0.07$ applied on each entangled qubit pair. 
    The correct outcome is given by $100$ in binary.}
    \label{fig:sum_1,1,1,1_histogram_p_07_dephasing}
\end{figure}

In the noisy setting, the correct outcome, $4$ or $100$ in binary, has the highest probability, followed by outcome $000$. 
The probability distribution seems to have some symmetry (cf. Section~\ref{sec:analytical_study}), hence Figure~\ref{fig:sum_1,1,1,1_polar_p_07_dephasing} shows the probability distribution in a polar plot. 
The second most frequent outcome is diametrically opposed to the correct one; 
and the next two most frequent outcomes are $\pi/2$ radians away from the highest probability and diametrically opposed to each other. 
\begin{figure}[H]
    \centering
    \includegraphics[width=0.4\textwidth]{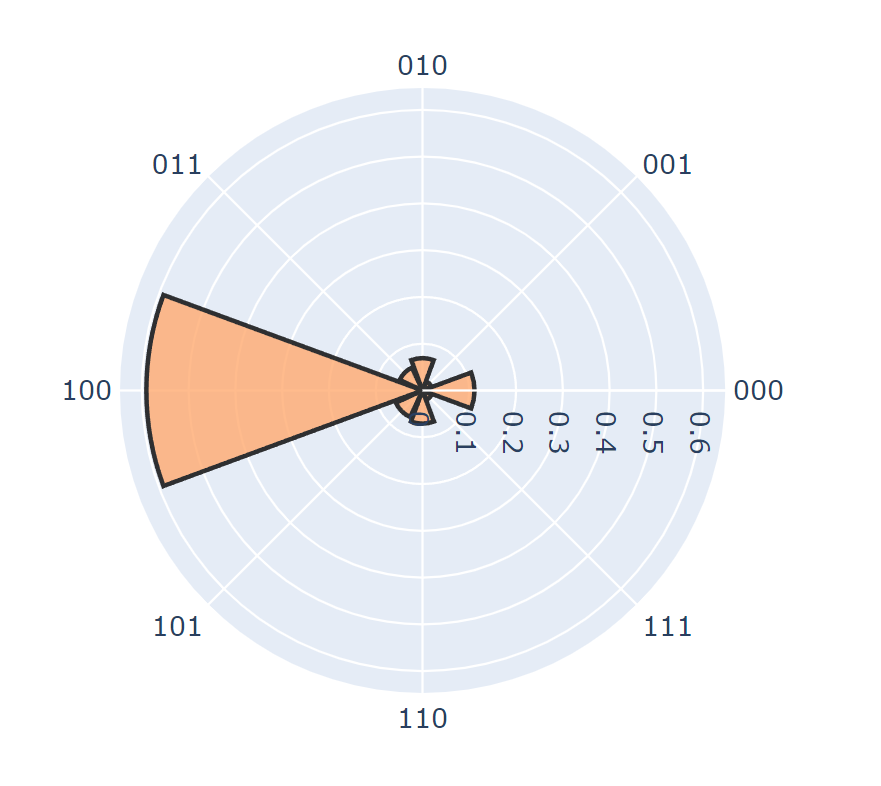}    
    \caption{Polar representation of the distribution shown in Figure~\ref{fig:sum_1,1,1,1_histogram_p_07_dephasing}.}
    \label{fig:sum_1,1,1,1_polar_p_07_dephasing}
\end{figure}

\begin{figure}[H]
    \centering
    \textbf{Four parties, dephasing noise}\par\medskip
    \subfigure[$p/2=0.05$]{\includegraphics[width=0.4\textwidth]{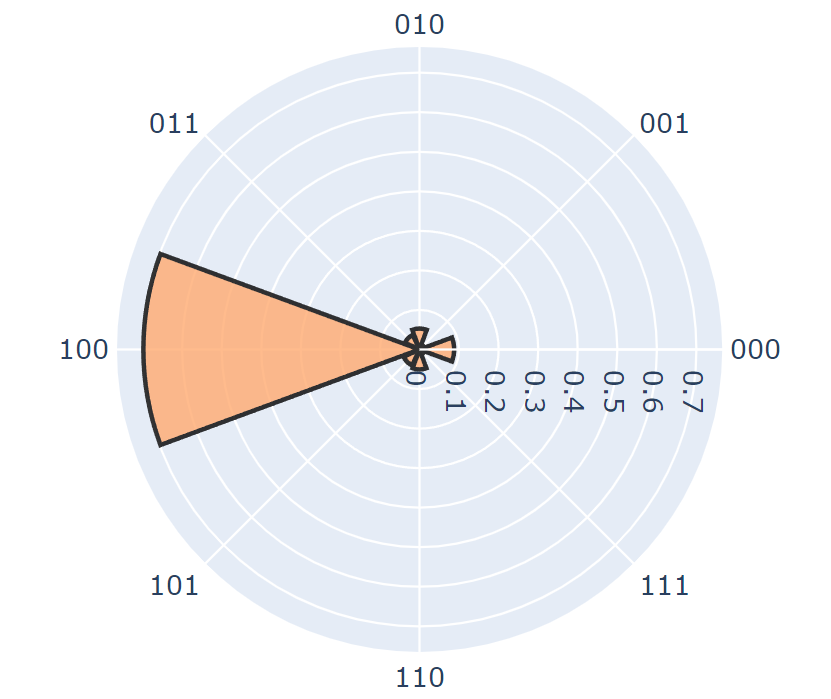}}
    \subfigure[$p/2=0.25$]{\includegraphics[width=0.4\textwidth]{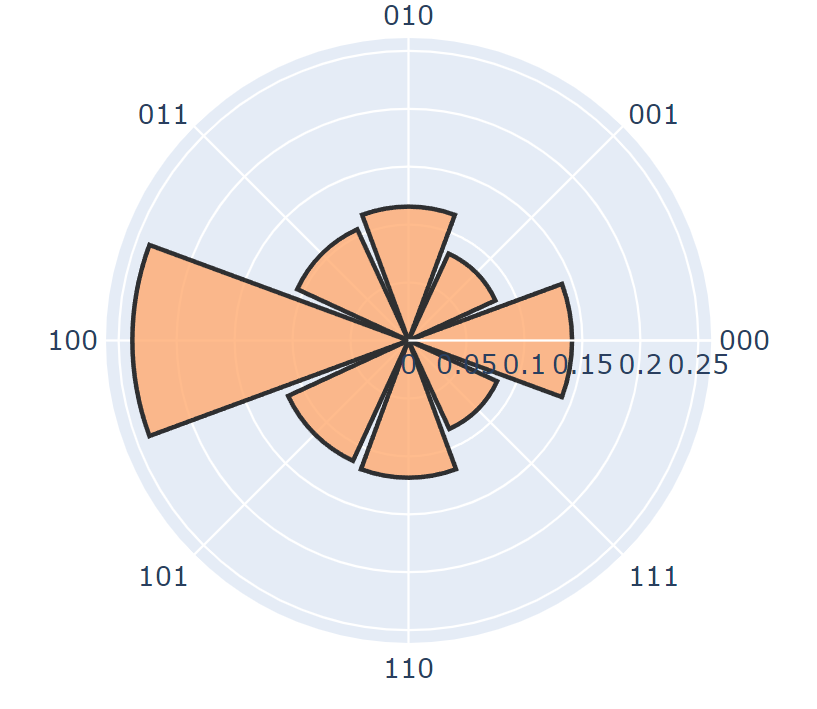}}
    \caption{Probability distribution for the case of four parties inputting $1$, with identical dephasing noise applied on each entangled qubit pair for varying noise levels, $p/2$, as indicated. 
    The correct outcome is given by $100$ in binary.}
    \label{sum_1111_dephasing}
\end{figure}

\begin{figure}[H]
    \centering
    \textbf{Two parties, dephasing noise}\par\medskip
    \subfigure[$p/2=0.05$]{\includegraphics[width=0.4\textwidth]{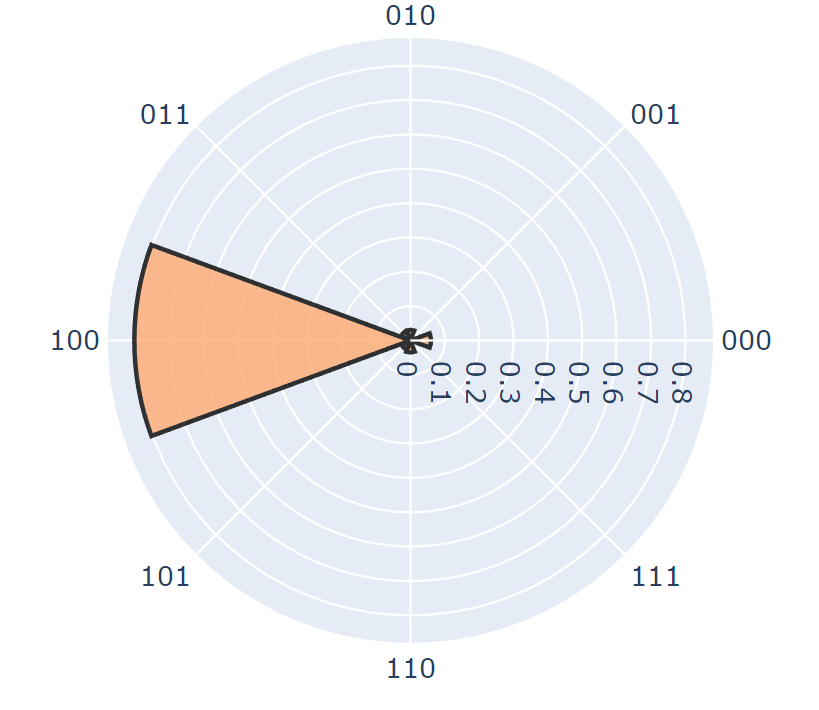}}
    \subfigure[$p/2=0.25$]{\includegraphics[width=0.4\textwidth]{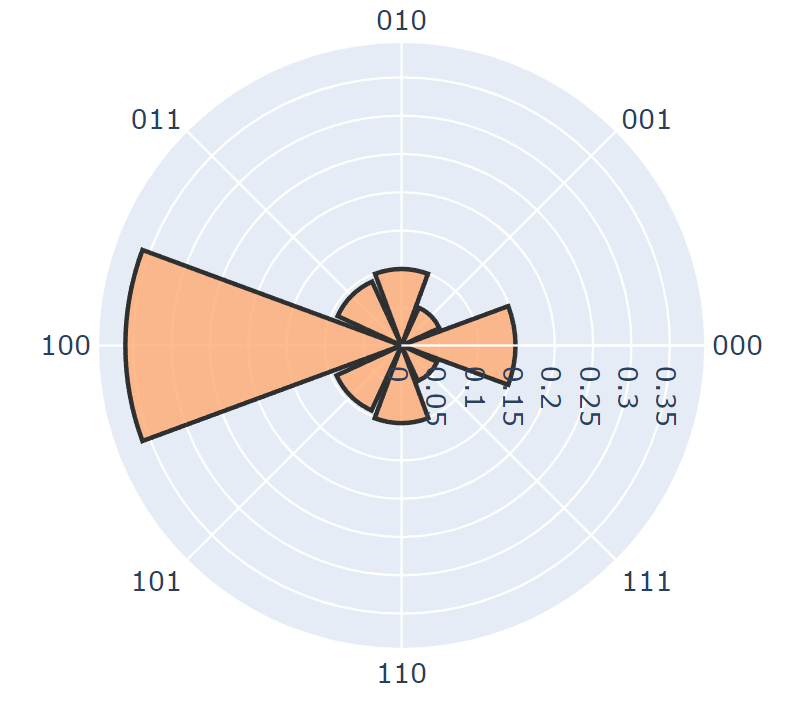}}
    \caption{Probability distribution for the case of two parties inputting $2$, with identical dephasing noise applied on each entangled qubit pair for varying noise levels, $p/2$, as indicated. 
    The correct outcome is given by $100$ in binary.}
    \label{sum_22_dephasing}
\end{figure}

\begin{figure}[H]
    \centering
    \textbf{Two parties, dephasing noise}\par\medskip
    \subfigure[$p/2=0.05$]{\includegraphics[width=0.4\textwidth]{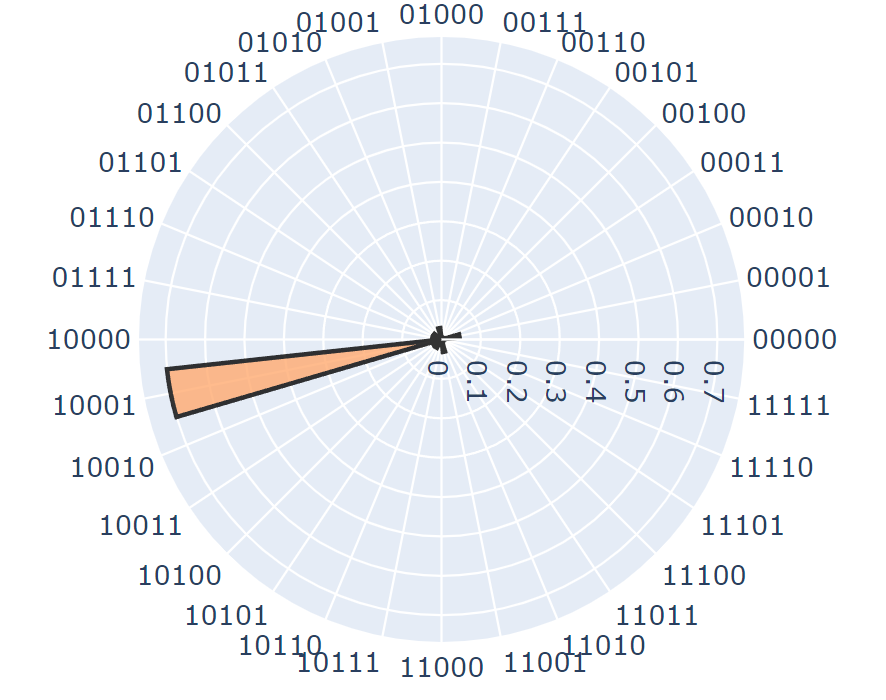}}
    \subfigure[$p/2=0.25$]{\includegraphics[width=0.4\textwidth]{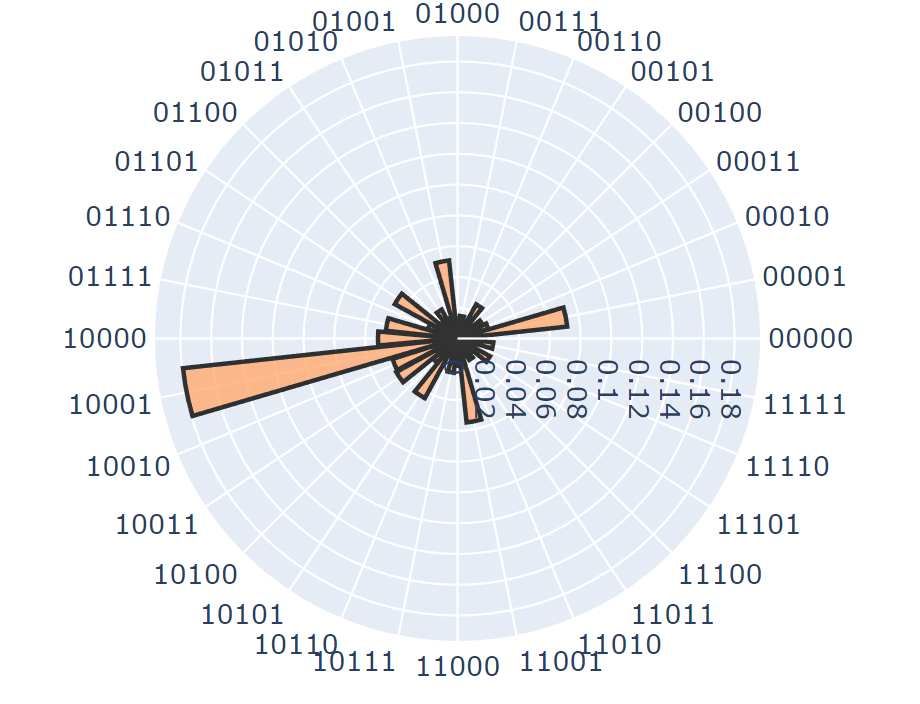}}
    \caption{Probability distribution for the case of two parties with inputs $10$ and $7$, respectively, with identical dephasing noise applied on each entangled qubit pair for varying noise levels, $p/2$, as indicated. 
    The correct outcome is given by $10001$ in binary.}
    \label{sum_107_dephasing}
\end{figure}

Interestingly, this symmetry emerges also for a different number of parties and for varying inputs. 
First, Figure~\ref{sum_1111_dephasing} shows the results for the same number of parties and inputs but with different levels of noise. 
Even though the output distribution approaches a uniform distribution with increasing noise levels $p$, the symmetry in the probability distribution is preserved.
Second, Figure~\ref{sum_22_dephasing} shows the results for a protocol run with two parties, both of them inputting $2$, for varying values of $p$.
Again, a similar noise pattern emerges, indicating that the noise pattern is independent of the number of parties involved. 
Finally, Figure~\ref{sum_107_dephasing} shows the results for a protocol run with two parties, where one of them inputs $10$ and the other $7$, again for varying values of $p$.
We again see similar symmetries appearing in the noise probability distribution, which indicates that the probability distribution is independent of the input values of the parties. 

\subsection{Depolarising noise}\label{sec:depolarising_noise}
We similarly performed the analysis for depolarising noise instead of dephasing noise. 
Interestingly, the same probability distributions were found as for dephasing noise, with the same symmetry patterns emerging. 
We therefore omitted the figures, as they give no additional information compared to the figures in the previous circuit. 
In the next section, we prove that indeed the probability distributions follow a specific pattern with symmetries, independent of the type of noise.

\section{Analytical study}\label{sec:analytical_study}
The probability distributions shown in the previous sections show some symmetry.
In this section we analyse this symmetry effect and show that, for fixed error probability, indeed the weighted Hamming distance with the correct output string determines the probability of being measured. 
We derive an expression for the probability distribution that applies to both dephasing and depolarising noise, and then discuss the intuition on the relation between the analytical expression and the observed pattern. Proofs of the results presented in this section can be found in Appendix \ref{sec:appendix}.

\subsection{Proof of probability distribution}
The probability distribution of the \textbf{DQA} under dephasing or depolarising noise follows a specific probability distribution. 
\begin{lemma}\label{lemma:noise}
    Under depolarising or dephasing noise, the server party state right before the application of the Inverse Quantum Fourier Tranform on a \textbf{DQA} for $n$ parties can be written as 
    \begin{equation}
    \rho  = \bigotimes_{s=0}^{n-1}\frac{1}{2}(\ket{0}\bra{0}+\ket{1}\bra{1}+ae^{-i\theta_s}\ket{0}\bra{1} + ae^{i\theta_s}\ket{1}\bra{0}),
    \label{server_state_noise}
\end{equation}
where $a=\prod_{i=0}^{n-1}(1-p_i)^2$ for dephasing noise and $a=\prod_{i=0}^{n-1}[1-(1-p_i)^2]$ for depolarising noise, with $p_i$ the noise parameter for party $i$.
\end{lemma}

Now, to characterise the output probability distribution of the distributed adder protocol, we need to look at the product of the $\Tilde{a}_i$ factors. 
As we consider the same error rates for every qubit, we define the fidelity parameter $\Tilde{a}$ as $\Tilde{a} = \prod_{i=0}^{n-1}\Tilde{a}_i$.
In particular, $\Tilde{a}=1$ corresponds to a noiseless GHZ-state, whereas $\Tilde{a}=0$ corresponds to a completely dephased or depolarised GHZ state. 

\begin{theorem}
Let $\{t^{(i)}\}_{i=1}^m$ be the inputs of $m$ different parties and let for each $s \in \{0,1,\ldots n-1\}$
\begin{equation}
    \theta_s = \frac{\pi}{2^{n-1-s}}\sum_{i=1}^m t^{(i)}= \frac{2\pi}{2^{n-s}}\sum_{i=1}^m t^{(i)}.
\end{equation}
Then, the $m$-player \textbf{DQA} protocol produces the output probability distribution such that for each potential output $x$
\begin{equation}
    P(x)=\frac{1}{2^n}\prod_{s=0}^{n-1}[1+ a \cos{(\theta_s-2\pi x/2^{n-s})}],
    \label{eq:main_result}
\end{equation}
with fidelity parameter $a\in[0,1]$ related to the depolarising or dephasing noise level. 
\end{theorem}

\subsection{Understanding the noisy distribution}
This section provides intuition for what the proven theoretical distribution in Equation~\eqref{eq:main_result} actually looks like and how it translates to the distribution observed in the simulations.
From the equation, it follows that the probability is maximised if all cosines evaluate to $1$, which happens precisely if the argument of the cosines is an integer multiple of $2\pi$. 
Setting $x=\Tilde{x}$, with $\Tilde{x}$ the correct outcome of the summation, this indeed maximises the probability. 
The resulting probability then equals 
\begin{equation*}
    P(\Tilde{x})=\left(\frac{1+a}{2}\right)^n.
\end{equation*}
Note that for a noiseless setting where $a=1$, this indeed yields a probability of 1.

Now, if $y\in \mathbb{Z}_{2^n}$ is a different outcome, $y$ can be written as $y=\Tilde{x} + z$ for some error $z\in \mathbb{Z}_{2^n}\backslash\{0\}$. 
The probability to observe $y$ then equals
\begin{equation}\label{probabilities}
    P(y)=\frac{1}{2^n}\prod_{s=1}^{n}[1+a\cos(2\pi z/2^{s})]
\end{equation}
Note that we relabeled the counter with respect to Equation~\eqref{noise_pattern}. 
For every $s$, we see a periodic behavior in $z$ resulting from the cosines. 
Combined, we get a complex periodic behavior in the probability distribution of the possible outcomes.

Specifically, this allows us to prove that the probability distribution is symmetric around $x$:
\begin{lemma}
    Let $z\le 2^{n-1}$, then $P(x + z) = P(x +(2^n-z))$.
    \label{lem:base_case}
\end{lemma}
By the previous lemma, information on noise strings $z\le 2^{n-1}$ gives sufficient information on all possible noise strings. 

In addition, this allows us to show the behaviour observed in Section 3 regarding the second and third most frequent outcomes:
\begin{lemma}
    For any integer $k\in\{1,\hdots,n-1\}$ and any error string $z\in \{1,2,\hdots 2^{k}\}$, we have that
    \begin{equation*}
        P(x+2^k)\geq P(x+z)
    \end{equation*}
\end{lemma}

This lemma yields indeed that the state diametrically opposite to the correct outcome has the second largest probability, the states $\pi/2$ radians away from the correct outcome have the third largest outcome, and so forth.

In addition, we see that the probabilities closer to the correct value are larger. To be more precise:
\begin{lemma}
    For any integer $k\in\{1,2,\hdots,n-1\}$ and error string $z\in \{1,2,\hdots, 2^{k}\}$, we have
    \begin{equation*}
        P(x + z)\ge P(x + (2^{k+1}-z))
    \end{equation*} 
\end{lemma}

Running the circuit multiple times gives samples from the probability distribution. 
It would be natural to try and use multiple circuit runs to increase the probability of obtaining the correct answer, for example by comparing the obtained distribution with the theoretical distribution derived above.
However, as for fixed fidelity parameter $a$ the probabilities are exponentially small in $n$, standard techniques using Chernoff bounds require an exponential number of samples to lower bound the success probability of retrieving $x$.

\section{Protocol without Trusted Server}
\label{sec:protocol_modifications}
Like most multiparty protocols, the \textbf{DQA} protocol requires a trusted third party. 
Although this is a common assumption in some classical multiparty computation protocols, it is unrealistic in practice. 
This section therefore introduces a modification to the protocol which eliminates the necessity of this reliable authority. 

For this modification to work, a certain primitive from cryptography is required, called \textit{(Treshold) Shamir's Secret Sharing}~\cite{Shamir:1979}.
First, Threshold Shamir's Secret Sharing is briefly explained, after which the modified protocol is proposed.

\subsection{Threshold Shamir's Secret Sharing}
Threshold Shamir's Secret Sharing (SSS) is a well-known classical protocol originally intended to distribute secrets between several entities, with the ability to reconstruct them later~\cite{Shamir:1979}. 

Suppose an agent desires to distribute a secret $X$ among $k$ parties.
Then the agent would choose a polynomial of order $t<k$ over a finite field $GF(q)$
\begin{equation}
    g(x) = X + a_1 x+ \ldots + a_{t-1} x^{t-1}
\end{equation}
with a prime number $q$ such that $X<q$. 
Next, the secret sharer would choose a set of $k$ different points $\{x_1,\ldots,x_k\}$ to evaluate the polynomial on and would send one polynomial evaluation $g(x_j)$ to each party. 
Now, any subset of $t$ parties can reconstruct the secret by simply performing polynomial interpolation using the polynomial evaluations that they received and then determining $g(0)=X$.
Note that at least $t$ parties are required to perform this interpolation, as the polynomial has degree $t-1$.
Importantly, the knowledge of $t-1$ shares $\{(x_i,g(x_i))\}$ does not provide any information about the secret.

\subsection{Protocol modification}
A key observation is that SSS can be utilized to perform multiparty summation in a secure manner by combining it with repeated usage of \textbf{DQA}. 
The idea is to perform multiple rounds of SSS. 
In each round, one of the summing parties acts as the server party of the \textbf{DQA}, and each of the other parties input one of their shares for the quantum protocol.
The party acting as server then receives the sum of these inputted shares at the end of the round. 
In this way, no shares are sent to other parties, but are only used as input for the quantum protocol.
No party therefore learns about the shares of other parties.
By astutely combining the shares of different parties, the summation result can be reconstructed for any subset of at least $t$ parties.
This results in the following protocol:

\textbf{NO-TP-ADDER (NTPA)}

Consider $m$ parties, each holding a number $X_i$. 
\begin{itemize}
    \item \textbf{Step 1:} the parties jointly agree on a sufficiently large prime $q$ and make $q$ public;
    \item \textbf{Step 2:} each party $i$ chooses a polynomial of order $t$ over the finite field $GF(q)$
    \begin{equation}
        g_i(x) = X_i + a_1^i x+ \ldots + a_{t-1}^i x^{t-1}
    \end{equation}
    where the coefficients are chosen at random but non-zero, except for $a_0^i$, which corresponds to the real input from the party ($a_0^i=X_i$);
    \item \textbf{Step 3:} In each round, a different party will act as the server party, which requires agreement on the order for the parties to act as server;
    \item \textbf{Step 4:} $m$ rounds of \textbf{DQA} are performed. 
    For each $r\in\{1,\ldots,m\}$:
    \begin{itemize}
        \item Party $r$ acts as server;
        \item Party $i$ inputs share $g_i(r)$;
        \item Party $r$ receives the partial summation $\sum_i g_i(r)$
    \end{itemize}
    At the end of all rounds, each party $r$ has received the partial summation $\sum_i g_i(r)$. It hence knows $G(r)$ for $G(x)=\sum_{i=1}^{m} g_i(x)$
    \item \textbf{Step 5:} Now, the summation result can be restored by having $t$ parties cooperate to share their intermediate results $G(r)$. As $G(x)$ has degree at most $t-1$, these $t$ evaluation points of $G(x)$ are hence sufficient to reconstruct $G(x)$, from which the summation result can be computed by evaluating $G(0)=\sum_{i=1}^{m} g_i(0)=\sum_{i=1}^m X_i$. Note that just like in the original SSS protocol, at most $t-1$ shares are not sufficient to conclude anything about $\sum_{i=1}^{m} X_i$. 
\end{itemize}

The advantage of this approach is that by having each party act as a server in one round, there is no agent that holds more power or knowledge than any other.
Thus, the necessity for a trusted third party is removed, making the protocol more suitable for real life situations. 

It is important to note that since the parties only know the intermediate shares $G(r)$, and they cannot reconstruct the sum on their own. 
Knowledge of at least $t$ shares is needed to recover the result of the summation.

\section{Conclusions and outlook}\label{sec:conclusions}

As current quantum hardware is noisy, and is expected to remain so for a while, it is important to study the impact that imperfect operations have on the fidelity of quantum protocols.
Multiple works on noisy quantum algorithms are available in literature. 
Similarly, a few works on distributed quantum computing are available. 
This work combines distributed noisy operations and analyses the effect of imperfect operations on the outcome.
We restricted ourselves to imperfect shared entangled states with perfect operations on the individual devices, as the errors within local devices are generally smaller than the ones seen on quantum communication. 

We considered a the practical implementation of the distributed multiparty quantum summation protocol~\cite{Neumann2022}.
The shared entangled states undergo depolarising or dephasing noise, two common noise models that allow for an analytic study of the behaviour of the protocol. 
The probability distributions corresponding to the final state of a noisy summation protocol execution under these noise models show a clear symmetric pattern, which was proved in the analytic study.
The magnitude of the probability on the erroneous states depends on the amount of noise affecting the execution of the protocol and how far the erroneous string is from the correct outcome.
Knowledge about this distribution and multiple samples can help point towards what the actual outcome may be. 
However, for constant noise parameters, an exponential number of samples is needed to actually determine the correct outcome. 

The protocol initially uses a trusted third party. 
In practice, this assumption is however unrealistic.
This need for a trusted party was removed by building upon the classical Shamir's Secret Sharing protocol.
As an added benefit, all parties automatically learn the outcome of the protocol. 
As the number of protocol rounds increases, the effect of noise is increased, diminishing the probability of obtaining the correct result of the summation.

Future work should address the effects of other sources of noise that are present in the protocol, such as imperfect local operations.
More importantly, a proper study of the effects that noise has on the security of the protocol needs to be done. 
In a perfect noiseless setting, the quantum no-cloning theorem ensures that no information is leaked to the outside world without corrupting the states that the parties share;
thus, the parties could detect the presence of an adversary and abort the protocol before inputting their secrets. 
However, the signature on the shared entangled states of the action of an eavesdropper is indistinguishable from noise. 
Hence, the parties cannot expose an eavesdropper just by checking quantum correlations via the shared entangled states. 
In this case, the amount of information that can be leaked and learned by an eavesdropper from the execution of the protocol should be bounded. 
Typically, these bounds depend on the amount of noise present in a protocol run, such that security can be guaranteed under certain noise conditions (see for example~\cite{Khabiboulline}).
As a first step, formal definitions of security, anonymity and privacy in the context of quantum multiparty summation must be established, similar to the contribution that Arapinis et al. had in the field of Quantum Electronic Voting~\cite{arapinis2021definitions}.

Although the conclusions from the present study are specific for the investigated protocol, most quantum multiparty protocols rely on the utilisation of entanglement at some stage. 
Thus, the methodology used here can inspire the analysis of noise robustness in similar protocols. 

\appendix
\setcounter{theorem}{0}
\section{Proofs of Results of Section \ref{sec:analytical_study}}\label{sec:appendix}
\begin{lemma}
Under depolarising or dephasing noise, the state of the server party right before the application of the Inverse Quantum Fourier Tranform on a \textbf{DQA} run for $n$ parties can be written as in \ref{server_state_noise};
where $a=\prod_{i=0}^{n-1}(1-p_i)^2$ for dephasing noise and $a=\prod_{i=0}^{n-1}[1-(1-p_i)^2]$ for depolarising noise, with $p_i$ the noise parameter for party $i$.
\end{lemma}
\begin{proof}
From the effect that a one qubit \textit{dephasing channel} has on an $n$-qubit $GHZ$ state
\begin{align*}
    \rho & \mapsto \Big(1-\frac{p_i}{2}\Big)\rho + \frac{p_i}{2}Z_i\rho Z_i \\
    & = \frac{1}{2}\bigg(\ket{0}^{\otimes n}\bra{0}^{\otimes n}+\ket{1}^{\otimes n}\bra{1}^{\otimes n} + (1-p_i) \Big(\ket{0}^{\otimes n}\bra{1}^{\otimes n} + \ket{1}^{\otimes n}\bra{0}^{\otimes n}\Big)\bigg),
\end{align*}
it can be shown that the application of a noisy non local CNOT between server party and $n$ parties takes their joint state to
\begin{equation*}
    \rho_{s,1,\ldots,n} \mapsto \frac{1}{2} \bigg(\ket{0}^{\otimes n+1}\bra{0}^{\otimes n+1}+\ket{1}^{\otimes n+1}\bra{1}^{\otimes n} + \prod_{i=0}^{n-1}(1-p_i) \Big(\ket{0}^{\otimes n+1}\bra{1}^{\otimes n+1} + \ket{1}^{\otimes n+1}\bra{0}^{\otimes n+1}\Big)\bigg).
\end{equation*} 
Then, the parties input their corresponding data through the $\mathcal{Z}$ rotations of an angle $\theta_j$. After that, a
second distributed CNOT under dephasing noise with parameter takes the state of the server to state in \ref{server_state_noise}, with $a=\prod_{i=0}^{n-1}(1-p_i)^2 $

For \textit{depolarising noise}, the same expression holds by replacing $\prod_{i=0}^{n-1}(1-p_i)$ by $1-(1-p_i)^2$.
Note that when dealing with more than $2$ qubits, applying an $n$-depolarizing channel gives a different result compared to applying a $1$-depolarizing channel on each qubit. 
Here we are considering $n$-qubit depolarizing channels.

Let $\rho_j$ be the state of qubit $j$ of the server. After entanglement generation and the first non local CNOT under depolarizing error with parameter $p_{i,1}$, the parties input their corresponding data through the $\mathcal{Z}$ rotations of an angle $\theta_j$. Then, a
second distributed CNOT under depolarizing noise with parameter $p_{i,2}$ takes the state to
\begin{align*}
    \rho_j &=(1-p_{i,1})\Big((1-p_{i,2}){\rho}'_j \otimes \ket{0}\bra{0}^{n}+ \frac{p_{i,1}}{2} \mathrm{I}_{n+1}\Big)+\frac{p_{i,2}}{2}\mathrm{I}_{n+1} \\
     &= \Bigg((1-p_{i,1})(1-p_{i,2}){\rho}'_j\otimes \ket{0}\bra{0}^{n} + \frac{p_{i,1}(1-p_{i,2})+p_{i,1}}{2} \mathrm{I}_{n+1}\Bigg)   \nonumber \\ 
\end{align*}
with $\rho'_j=\frac{1}{2}\Big(\ket{0}\bra{0}+\ket{1}\bra{1}+ e^{-i\theta_j}\ket{0}\bra{1} + e^{i\theta_j}\ket{1}\bra{0}\Big)$.
Taking $p_{i,1}=p_{i,2}=p_i$, setting $a = 1-(1-p)^2$ and tracing out the degrees of freedom of the non server parties, we can write
\begin{equation}
    \rho''_j= \mathbf{Tr}_{1\ldots n}[\rho_j]  = \frac{1}{2}\Big(\ket{0}\bra{0}+\ket{1}\bra{1}+ a e^{-i\theta_j}\ket{0}\bra{1} + a e^{i\theta_j}\ket{1}\bra{0}\Big) .
\end{equation}
\end{proof}

\begin{theorem}
The $m$-player \textbf{DQA} protocol produces the output probability distribution such that for each potential output $x$
\begin{equation}
    P(x)=\frac{1}{2^n}\prod_{s=0}^{n-1}[1+ a \cos{(\theta_s-2\pi x/2^{n-s})}],
\end{equation}
with fidelity parameter $a\in[0,1]$ related to the depolarising or dephasing noise level in the shared GHZ states. 
\end{theorem}

\begin{proof}
The state just before the final inverse quantum Fourier transform is given by 
\begin{equation}
    \ket{\Psi} = \bigotimes_{s=0}^{n-1} \frac{1}{\sqrt{2}}(\ket{0}+e^{i\theta_s}\ket{1})= \bigotimes_{s=0}^{n-1} \frac{1}{\sqrt{2}}\left(\sum_{j_s=0,1}e^{i\theta_s j_s}\ket{j_s}\right),
\end{equation}
which simplifies to
\begin{equation}
    \ket{\Psi}=\frac{1}{\sqrt{2^n}}\sum_{j_=0}^{2^n-1}e^{i\sum_{s=0}^{n-1} \theta_s j_s}\ket{j_{n-1}\ldots j_0}.
\end{equation}

From Lemma~\ref{lemma:noise}, the presence of dephasing or depolarising noise on the quantum edges mixes the state of the server, such that the density matrix takes the form
\begin{equation}
    \rho  = \bigotimes_{s=0}^{n-1}\frac{1}{2}(\ket{0}\bra{0}+\ket{1}\bra{1}+ae^{-i\theta_s}\ket{0}\bra{1} + ae^{i\theta_s}\ket{1}\bra{0})
\end{equation}
which can be written as
\begin{equation}
    \rho = \frac{1}{2^n} \sum_{j=0}^{2^n-1} \sum_{k=0}^{2^n-1} a^{\sum_s|k_s-j_s|} e^{i\sum_s(\theta_s j_s-\theta_s k_s)} \ket{j_{n-1}\ldots j_0} \bra{k_{n-1}\ldots k_0},
    \label{eq:density_matrix:Depolarising_noise}
\end{equation}
where $a=\Tilde{a}^2$, as the distributed CNOT gates are performed twice, before and after the rotations.

The last step of the protocol consists of applying the inverse quantum Fourier transform.
It maps the state in Equation~\eqref{eq:density_matrix:Depolarising_noise} to
\begin{align}
& \mathbf{IQFT}_n\rho \mathbf{QFT}_n \nonumber \\
& \quad = \frac{1}{2^n} \sum_{j,k=0}^{2^n-1} a^{\sum_s|k_s-j_s|} e^{i\sum_s(\theta_s j_s-\theta_s k_s)} \left( \mathbf{IQFT}_n \ket{j_{n-1}\ldots j_0} \right)\left(\mathbf{IQFT}_n \ket{k_{n-1}\ldots k_0}\right)^\dagger\nonumber \\
& \quad = \frac{1}{2^{2n}} \sum_{x,y=0}^{2^n-1} \sum_{j,k=0}^{2^n-1} a^{\sum_s|k_s-j_s|} e^{i\sum_s\left(\theta_s-\frac{\pi}{2^s} x\right)j_s}e^{-i\sum_s\left(\theta_s - \frac{\pi}{2^s} y\right)k_s }\ket{x_{n-1}\ldots x_0} \bra{y_{n-1}\ldots y_0}.
\end{align}
As the probability of measuring a computational basis state corresponds to the corresponding diagonal element of the density matrix, we obtain the probability $P(x)$ by setting $x=y$:
\begin{align}
    P(x) & = \frac{1}{2^{2n}} \sum_{j,k=0}^{2^n-1} a^{\sum_s|k_s-j_s|} e^{i\sum_s\left(\theta_s-\frac{2\pi}{2^{n-s}}x\right)j_s}e^{-i\sum_s\left(\theta_s - \frac{2\pi}{2^{n-s}}x\right)k_s } \nonumber \\
    & = \frac{1}{2^{2n}} \prod_{s=0}^{n-1} \sum_{j_s,k_s=0,1} a^{|k_s-j_s|} e^{i\left(\theta_s-\frac{2\pi}{2^{n-s}}x\right)j_s}e^{-i\left(\theta_s - \frac{2\pi}{2^{n-s}}x\right)k_s } \nonumber \\
    & = \frac{1}{2^{2n}} \prod_{s=0}^{n-1}  \left(1 + a e^{i\left(\theta_s-\frac{2\pi}{2^{n-s}}x\right)} +a e^{-i\left(\theta_s-\frac{2\pi}{2^{n-s}}x\right)} +1 \right) \nonumber \\ 
    & = \frac{1}{2^{n}} \prod_{s=0}^{n-1}  \left[1 + a \cos{\left(\theta_s-\frac{2\pi}{2^{n-s}}x\right)} \right], 
    \label{noise_pattern}
\end{align}
which completes the proof. 
\end{proof}

\begin{lemma}
    Let $z\le 2^{n-1}$, then $P(x + z) = P(x +(2^n-z))$.
\end{lemma}
\begin{proof}
    By Equation~\eqref{probabilities} we have
    \begin{align*}
        P(x + (2^n-z)) & = \frac{1}{2^n}\prod_{s=1}^{n}[1+a\cos(2\pi (2^n-z)/2^{s})] \\
        &= \frac{1}{2^n}\prod_{s=1}^{n}[1+a\cos(2\pi 2^{n-s} - 2\pi z/2^s)] \\
        & = \frac{1}{2^n}\prod_{s=1}^{n}[1+a\cos(2\pi z/2^{s})] 
        = P(x + z).
    \end{align*}
    Here we used the identity $\cos(2\pi t + x) = \cos(x)$ for any integer $t$. 
\end{proof}

\begin{lemma}
    For any integer $k\in\{1,\hdots,n-1\}$ and any error string $z\in \{1,2,\hdots 2^{k}\}$, we have that
    $P(x+2^k)\geq P(x+z)$
\end{lemma}
\begin{proof}
    Write $z=2^t m$ with $m$ odd and $0\leq t \leq k$.
    By expanding Equation~\eqref{probabilities}, we get
    \begin{align*}
        P(x + 2^k) & = \frac{1}{2^n}\prod_{s=1}^{n}[1+a\cos(2\pi 2^k/2^{s})] 
        = \frac{1}{2^n}\prod_{s=1}^{k}[1+a\cos(2\pi 2^k/2^{s})]\prod_{s=k+1}^{n}[1+a\cos(2\pi 2^k/2^{s})] \\
        & = \frac{1}{2^n}(1+a)^{k}\prod_{s=k+1}^{n}[1+a\cos(2\pi 2^t/2^{s-k+t})].
    \end{align*}
    The cosines in the first product term all evaluate to $1$. 
    In the second product term, we rewrote the fraction in each of the cosines.
    Now, see that for each $s \in\{k+1, \hdots, n\}$, it holds that
    \begin{equation*}
        \cos(2\pi 2^t/2^{s-k+t})=\cos(2\pi /2^{s-k}) \geq \cos(2\pi m/2^{s-k}) = \cos(2\pi 2^t m/2^{s-k+t}).
    \end{equation*}
    The inequality follows as $m/2^{s-k}$ is an odd multiple of $1/2^{s-k}$ and hence will be at least as far from $0$ on the unit circle as $1/2^{s-k}$. 
    Applying this inequality to each of the terms of the second product, we obtain
    \begin{align*}
        P(x + 2^k) & \geq \frac{1}{2^n}(1+a)^{k}\prod_{s=k+1}^{n}[1+a\cos(2\pi z/2^{s-k+t})] 
        = \frac{1}{2^n}(1+a)^{t}\prod_{s=t+1}^{n-k+t}[1+a\cos(2\pi z/2^{s})](1+a)^{k-t} \\
        &\geq \frac{1}{2^n}\prod_{s=1}^{t}[1+a\cos(2\pi z/2^{s})]\prod_{s=t+1}^{n-k+t}[1+a\cos(2\pi z/2^{s})]\prod_{s=n-k+t+1}^{n}[1+a\cos(2\pi z/2^{s})] \\
        &= P(x+z)
    \end{align*}
    In the second line we relabeled the product index and reordered the $(1+a)$ terms. 
    In the last inequality we used that each cosine is at most $1$, so adding these terms can only make the term smaller. 
\end{proof}

\begin{lemma}
    For any integer $k\in\{1,2,\hdots,n-1\}$ and error string $z\in \{1,2,\hdots, 2^{k}\}$, we have 
    $P(x + z)\ge P(x + (2^{k+1}-z))$
\end{lemma}
\begin{proof}
    Expanding Equation~\eqref{probabilities}, we see
    \begin{align*}
        P(x + (2^{k+1}-z)) & = \frac{1}{2^n}\prod_{s=1}^{n}[1+a\cos(2\pi (2^{k+1}-z)/2^{s})] \\
        & = \frac{1}{2^n}\prod_{s=1}^{n-k}[1+a\cos(2\pi (2^{k+1}-z)/2^{s})]\prod_{s=n-k+1}^{n}[1+a\cos(2\pi z/2^{s})].
    \end{align*}
    The second equality followed from the identity $\cos(2\pi t-y)=\cos(y)$ for any integer $t$. 

    Next, note that $(2^{k+1}-z)\ge z$ and that $(2^{k+1}-z)/2^{s}) \le \frac{1}{2}$ for $s\le n-k$.
    As the cosine is decreasing on the $[0,\frac{1}{2}]$-interval it follows that 
    \begin{align*}
        P(x + (2^{k+1}-z)) & = \frac{1}{2^n}\prod_{s=1}^{n-k}[1+a\cos(2\pi (2^{k+1}-z)/2^{s})]\prod_{s=n-k+1}^{n}[1+a\cos(2\pi z/2^{s})] \\
        & \le \frac{1}{2^n}\prod_{s=1}^{n-k}[1+a\cos(2\pi z/2^{s})]\prod_{s=n-k+1}^{n}[1+a\cos(2\pi z/2^{s})] \\
        & = P(x + z).
    \end{align*}
\end{proof}